\documentclass[10pt, a4paper]{article}   %, dvips

\usepackage{epsfig}
\usepackage{epsf}
\usepackage{graphicx}
\usepackage{amssymb}
\usepackage{mathrsfs}
\usepackage{amsmath}

\begin{document}

\title{Robust Adaptive Control Charts}
\author{\large {Gejza Dohnal, Czech Technical University  in Prague}}
\date{}  % No date.
\maketitle

\begin{abstract}
In statistical process control, procedures are applied that require relatively strict conditions for their use. If such assumptions are violated, these methods become inefficient, leading to increased incidence of false signals. Therefore, a robust version of control charts is sought  to be less sensitive with respect to a breach of normality and independence in measurements. Robust control charts, however, usually increase the delay in the detection of assignable causes. This negative effect can, to some extent, be removed with the aid of an adaptive approach. 

\bigskip
Key words: statistical process control, sequential detection scheme, control chart,  robustness, adaptive control chart, ARL, CUSUM, EWMA.

\end{abstract}

\bigskip
\section{Introduction}
 
\medskip
Although Shewhart type control charts work in many situations, the prerequisites for their use are hardly ever exactly met.  In practice, the most commonly observed phenomenon is a violation of normality in the observed values. 
It often becomes evident due to the heavy tails when multiple observed measurements lie below the mean of the expected normal distribution. 
A very common phenomenon is the skewness of data. In this case, deviations from the mean value frequently occur that are biased to one side, and the distribution becomes asymmetrical. Due to the measuring devices and human error,  outliers occur, i.e., values significantly distant from the mean. Typically, this includes errors arising from the wrong decimal point when writing the measured values or zero measurement due to a failure in reading and writing of the measured values. In some cases, the measured value is, in addition to purely random factors, affected by other inherent or unrecognized systematic effects and the resulting distribution of values is a mixture of several distributions.

These and other deviations from the exact normal distribution usually lead to an increase in the frequency of false signals. This phenomenon is undesirable for several reasons -- it:

-- increases the costs of control,

-- increases the variability in the controlled process,

-- reduces the credibility of the control chart and is likely to lead to its merely formal use,
\phantom{mnn} and less attention is consequently devoted to the actual signals.

\medskip
Another phenomenon, leading to an increased frequency of false alarms, is a violation of independence in the measurements over time. It may, however, to some extent eliminate an appropriately chosen model (e.g., ARMA, ARIMA) and the observed characteristics indicate deviations from this model \cite{montgomery},\cite{amin},\cite{jensen}. 
Eliminating the influence of outliers or heavy tails is much more complicated. In such cases, we are looking for a robust version of control charts \cite{rocke}.

The effort to increase robustness of the control chart usually leads to two effects that occur simultaneously:
\begin {itemize}\setlength\itemsep{-1mm}
\item reduction in the frequency of false signals;
\item the resulting control chart becomes less sensitive to the behavior change of the process, in consequence of an assignable cause
\end {itemize}

\noindent The second effect can be just as annoying as the high frequency of false signals. Long delays in the signal for the emergence of assignable causes may lead to 

\smallskip
\noindent-- cost increases due to poor production when the process is in an out-of-control 
\phantom{n}  state

\noindent-- lower confidence in the effectiveness of the control chart

\smallskip
\noindent This unpleasant phenomenon can be partially eliminated using the adaptive approach to the sequential detection scheme \cite{tagaras},\cite{woodal},\cite{zimmer},\cite{tsung}.

\medskip
In Section 2 we will review some robust proposals based on robust estimates of control characteristics. Section 3 gives some examples of adaptive approaches, and designs a new adaptive scheme.

%%%%%%%%%%%%%%%%%%%%%%%%%%%%%%%%%%%%%%%%%%%%%%%%%%%%
\section{Robust control charts}
 
 \medskip
The robustness of control charts is usually understood as a control chart being less sensitive to violated assumptions of normality in the  observed characteristics. 
The robustness of control charts is usually increased in three ways:
\begin {itemize}\setlength\itemsep{-1mm}
\item adjusting control limits \cite{quesen},\cite{figueir},
\item using robust estimates of controlled characteristics, such as using the median estimate for the position or inter-quartile range to estimate variability \cite{jensen}, \cite{rocke}
\item using nonparametric tests in a sequential detection scheme \cite{lehman},\cite{chakra},\cite{baker} 
\end {itemize}

In this paper we will show results of applying the adaptive scheme on a robust CUSUM chart. We assume the following control chart based on the CUSUM statistics:
\begin{eqnarray*}
\tilde{C}^{+}_{n+1}= \max\big[ 0, \tilde{C}^{+}_{n}-(\mu_0+\delta_0)+\tilde{X}_n\big] \\
\tilde{C}^{-}_{n+1}= \max\big[ 0, \tilde{C}^{-}_{n}+(\mu_0+\delta_0)-\tilde{X}_n\big] 
\end{eqnarray*}
where $\mu_0$ is the desired process mean, $\delta_0$ is the threshold for shift detection, and $\tilde{X}_n$ is a median of the $n$-th sample. The signal will be sent out whenever $C^{+}_n \geq L$. This describes the robust version of a chart denoted as CUSUM $\tilde{X}$ in the text that follows (see, e.g., \cite{reynolds}).

%%%%%%%%%%%%%%%%%%%%%%%%%%%%%%%%%%%%%%%%%%%%%%%%%%%
\section{Adaptive control charts}

\medskip

An adaptive sequential detection scheme (SDS) responds to the current state of the process, estimated from the results of measurements made during the inspection. According to current information on the process, we choose parameters for the next SDS period (until the next inspection). A slow response of an adaptive procedure can, to a certain extent, be remedied using a robust control chart. An adaptive SDS is primarily intended to shorten the delay of the control chart.

\smallskip
Moreover, obtaining data in Phase I of SDS (see \cite{montgomery}) for estimation of parameters to reach an acceptable level is often too costly. An insufficient number of samples in Phase I can lead to large uncertainties in parameter estimation. An alternative is to improve estimation accuracy by using samples collected in Phase II using an adaptive approach. This leads us to adaptive control charts.
   
\medskip
In this context, we distinguish between two groups of parameters: sample parameters and design parameters of the chart. 
The sample parameters mainly include the time between inspections $h$ and sample sizes $n$ to be inspected. Shortening the time between inspections can significantly reduce the average delay of a control chart. Moreover, if the probability of failure (assignable cause) increases with time, long intervals can be set between inspections at the beginning of control and be further reduced with an increasing probability of failure. This can save considerable costs of control on the one hand, and shorten the response time to the emergence of assignable causes on the other hand. Increasing the number of measurements during the inspections in the event that we suspect the process to be out of statistical control, we refine our estimates of observed characteristics and can reduce the probability of false signals, while simultaneously increasing the likelihood of detecting faults in the process.

A disadvantage of sample parameters is that a change during the  process run requires organizational interventions in the work of operators. In contrast, design parameters, which include control limits, the threshold value in a CUSUM chart, weight $\lambda $ of an EWMA chart and others, can be changed independent of the operator, who takes samples and measurements. These parameter changes are easily accomplished as long as SDS is implemented with the aid of computers \cite{tsung}.

\bigskip
\subsection{Examples of charts with adaptive design parameters}

\medskip
\subsubsection{CUSUM adaptive control chart}

Sparks in \cite{sparks} suggested the following adaptive CUSUM control scheme  with an adaptive reference parameter:
$$
C_t=\max\big[0,\ C_{t-1}+(x_t-\delta_t/2)/h(\delta_t)\big]
$$
where $h(\delta_t)$ is a function which maintains a constant control limit. The shift magnitude  $\delta_t$  is updated on line using an EWMA-type equation 
$$ \delta_t=\max\big(wx_{t-1}+(1-w)\delta_{t-1}, \ \delta_{\min}\big).
$$
(suggested  $\delta_{\min}$  is 0.5 for detecting smaller shifts, 1.0 for detecting shifts larger than 1.0)

\medskip
\subsubsection{EWMA adaptive control chart} 

Capizzi and Masarotto in \cite{capizi} introduced an adaptive EWMA procedure with an adaptive smoothing parameter $w(e_t)$:
$$
Z_t=\big(1-w(e_t)\big)Z_{t-1}+w(e_t)x_t
$$
where $e_t=x_t-Z_t$. For small values of  $e_t$, $w(e_t)$ becomes relatively small, while for large $e_t$ the value of  $w(e_t)$ is enlarged accordingly. 

\subsubsection{Zone adaptive procedure with adaptive control limits}

The control chart of a zone type respects, in a certain manner, the supplementary run rules introduced by Western Electric in 1956. 
Suppose a zone control chart $\bar{X}$ with ten zones, defined by a central line $CL$ and four limits $k_1, k_2, k_3$ and $k_4$, which symmetrically space out the 
diagram into five zones below $CL$ (zones $-Z_0, -Z_1, -Z_2, -Z_3, -Z_4$) and five zones above $CL$ (zones $Z_0, Z_1, Z_2, Z_3, Z_4$ are on the left in Figure 1). Scores are assigned to points falling into each zone according to the right column in Figure 1. The function of a classical zone diagram is described in literature \cite{davis}, \cite{dohnal1}.  

\bigskip
\begin{center}
\begin{picture}(240,110)\setlength{\unitlength}{0.5mm}
\put(10,40){\line(1,0){160}} \put(-1,38){CL}
\put(10,80){\line(6,0){160}} 
\multiput(10,30)(4,0){40}{\line(1,0){2}} 
\multiput(10,20)(4,0){40}{\line(1,0){2}}
\multiput(10,10)(4,0){40}{\line(1,0){2}}
\multiput(10,50)(4,0){40}{\line(1,0){2}}
\multiput(10,60)(4,0){40}{\line(1,0){2}}
\multiput(10,70)(4,0){40}{\line(1,0){2}}
\put(10,0){\line(1,0){160}} 
\put(2,48){$k_1$}\put(2,58){$k_2$}\put(2,68){$k_3$}\put(2,78){$k_4$}
\put(-4,28){$-k_1$}\put(-4,18){$-k_2$}\put(-4,8){$-k_3$}\put(-4,-2){$-k_4$}
\put(139,42){$Z_0$}\put(139,52){$Z_1$}\put(139,62){$Z_2$}\put(139,72){$Z_3$}\put(120,82){zone: $Z_4$, score: 8}
\put(133.6,32){$-Z_0$}\put(133.6,22){$-Z_1$}\put(133.6,12){$-Z_2$}\put(133.6,2){$-Z_3$}\put(133.6,-8){$-Z_4$}
\put(169,42){0}\put(169,52){1}\put(169,62){2}\put(169,72){4}
\put(169,32){0}\put(169,22){1}\put(169,12){2}\put(169,2){4}\put(169,-8){8}
\put(10,-5){\line(0,1){90}} 
\put(25,39){\line(0,1){2}} \put(22.5,43){*} 
\put(15,39){\line(0,1){2}} \put(12.5,48){*}
\put(35,39){\line(0,1){2}}\put(32.5,20){*}
\put(45,39){\line(0,1){2}}\put(42.5,50){*}
\put(55,39){\line(0,1){2}}\put(52.5,39){*}
\put(65,39){\line(0,1){2}}\put(62.5,52){*}
\put(75,39){\line(0,1){2}}\put(72.5,28){*}
\put(85,39){\line(0,1){2}}\put(82.5,44){*}
\put(95,39){\line(0,1){2}}\put(92.5,63){*}
\put(105,39){\line(0,1){2}}\put(102.5,40){*}
\put(115,39){\line(0,1){2}}\put(112.5,58){*}
\multiput(124,38)(0,4){8}{\line(0,1){1}}
\put(122.5,68){*}
\put(122,32){$T$}
\end{picture}

\bigskip
Figure 1: Zone control chart.  \end{center}

\medskip
Here we introduce a new adaptive version of the zone chart. 

\noindent The decision rule is based on exceeding the lower or upper control limits (LCL, UCL),the same as in the Shewhart-type chart. These limits will be changed in each inspection step.
For each subgroup average, we determine into which zone $Z_k$ the value falls and look up the shrinkage parameter $s_k$ given for that zone. The control process starts with $(LCL_0, UCL_0)$. 
At each inspection $n$, if the subgroup average $\bar{X}_n$ is within the current control limits $(LCL_{n-1},UCL_{n-1})$, we will compute a new tuple of control limits $(LCL_n, UCL_n)$ in the following way:
\begin{eqnarray*}
(\bar{X}_n < 0) \wedge (\bar{X}_n  \in -Z_k) \Rightarrow  (LCL_n, UCL_n)=\big(L_n(LCL_{n-1}, s_k), UCL_0\big), \\
(\bar{X}_n \geq 0) \wedge (\bar{X}_n  \in Z_k) \Rightarrow  (LCL_n, UCL_n)=\big(LCL_0, L_n(UCL_{n-1}, s_k)\big),
\end{eqnarray*}
where the shrinkage function is defined as follows:
$$
L_n(LCL,s) = LCL + s\sigma_{\bar{X}}, \ \ \text{and} \ \  L_n(UCL,s) = UCL - s\sigma_{\bar{X}}.
$$

\noindent If the current subgroup average is on the same side of the central line as the preceding subgroup average, then the corresponding control limit will be shrunk, while the opposite control limit will be set to its maximal absolute value. As more subsequent observations are on the same side of CL,  the corresponding control limit will get close to CL and the probability of signal will increase. An example of such a control chart is shown in Figure 2.

\setcounter{figure}{1}
\begin{figure}[htb]
  \begin{center}
    \includegraphics[width=3in]{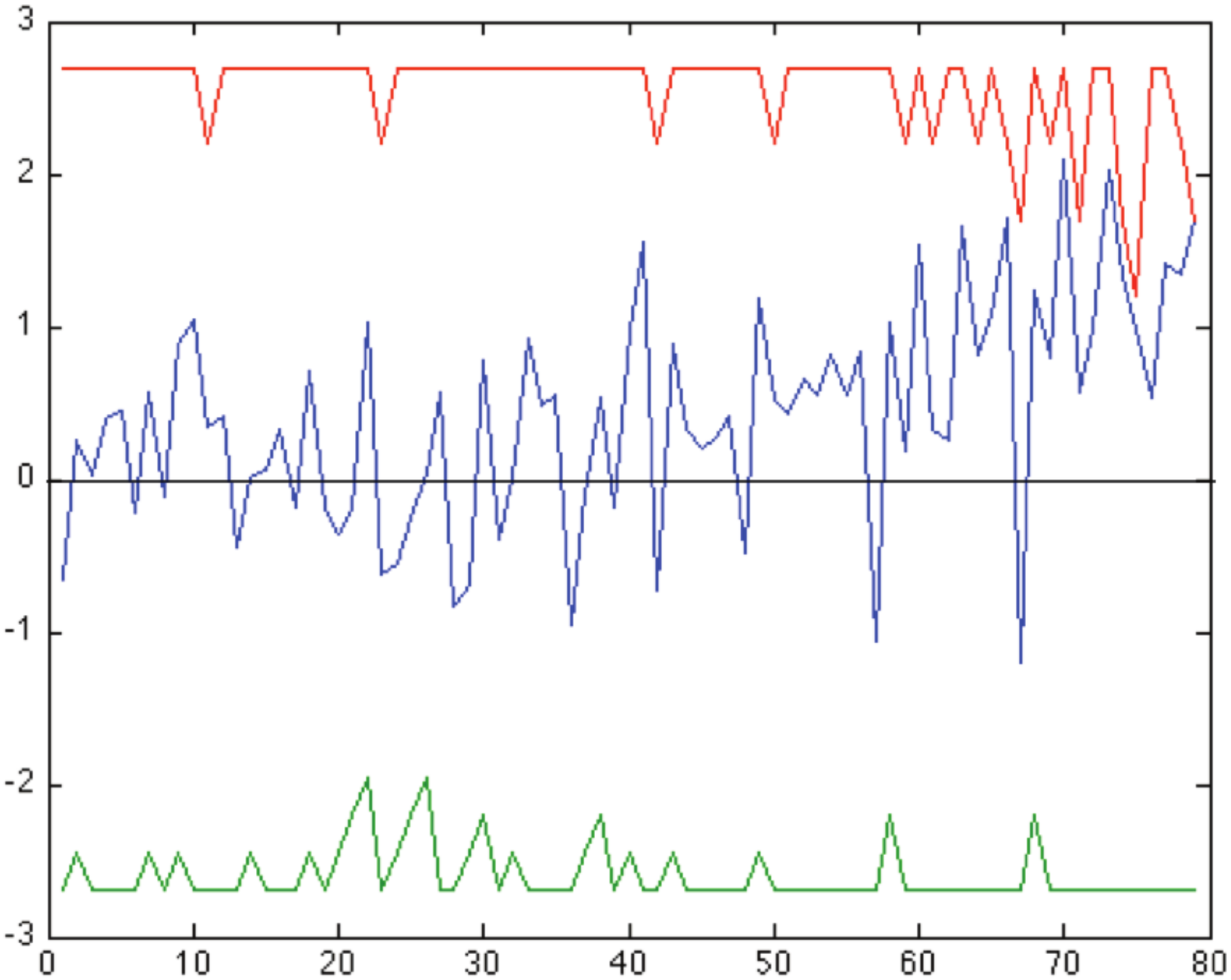}
  \caption{Detected shift.}  
  \end{center}
\end{figure}\label{fig2}

%%%%%%%%%%%%%%%%%%%%%%%%%%%%%%%%%%%%%%%%%%%%%%%%%%%%
\section{Robust adaptive control charts}

Let a control chart CUSUM $\tilde{X}$ be given which was introduced in Section 2. 
Suppose that parameters $\delta_0, \mu_0$ and $L$ for the process at in-control state were set within the design phase. 
When we apply to statistics $C^{+}_n, \ n=0,1,\dots$ the zone adaptive scheme described in the previous section, we obtain a robust adaptive control chart, which we shall denote as Ad-CUSUM $\tilde{X}$. The considered zone adaptive procedure is characterized by zone limits $\{k_0, k_1, k_2\}, k_2=L$ and shrinkage parameters $\{s_0, s_1, s_2\}$.  In Figure 3 we can see two possible runs of these charts.

\setcounter{figure}{2}
\begin{figure}[htb]
  \begin{center}
    \includegraphics[width=5in]{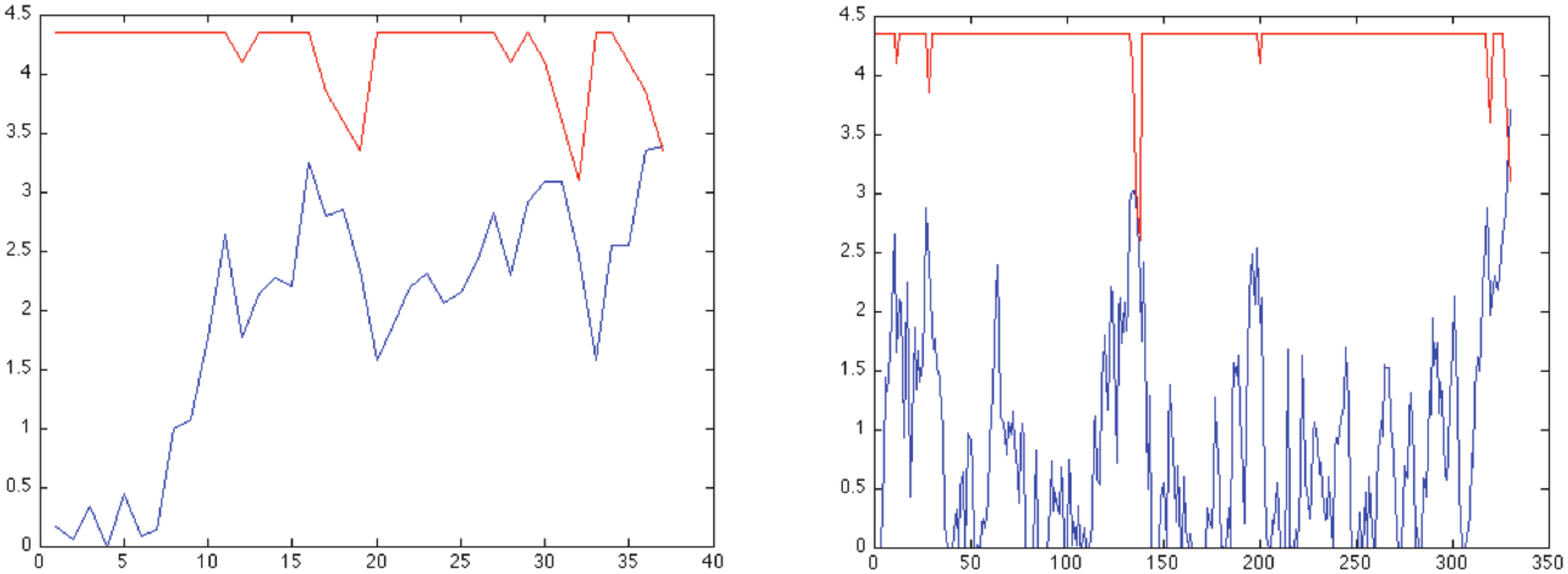}
  \caption{Detected shift.}  
  \end{center}
\end{figure}\label{fig3}

For a comparison of the listed diagrams' properties, we performed a numerical study in which we simulated data from both normal distribution $N(0,1)$ and a contaminated distribution, which means that observations $(100-\theta)\%$ come from the $N(\delta,1)$ normal distribution and $\theta\%$ come from the $N(\delta, \sigma^2)$ normal distribution, $\delta \geq 0$.  

\subsection{Numerical results}

The comparison of various control charts is based on ARL evaluation for both normal and contaminated data. The following settings were used:
\begin{enumerate}
\setlength\itemsep{-1.5mm}
\item The normal data are from the $N(\delta, 1)$ normal distribution.
\item The contaminated data consists of $94\%$ from the $N(\delta, 1)$ and $6\%$ from the $N(\delta, 6.25)$ normal distribution.
\item We simulated 10,000 runs with sample size $n=5$ for $\delta = 0.0, 0.1, 0.3,$ $0.5,$ $0.7, 1.0, 1.5$.
\item The following parameters were used for various control charts:
\begin{enumerate}
      \item Shewhart $\bar{X}$: $CL=0, LCL=-3.09, UCL=3.09$  
      \item Shewhart $\tilde{X}$: $CL=0, LCL=-3.128, UCL=3.128$  
      \item CUSUM $\bar{X}$: $\delta_0=0.15, L=4.001$  
      \item CUSUM $\tilde{X}$: $\delta_0=0.15, L=4.344$  
      \item EWMA $\bar{X}$: $\lambda=0.1, L=2.835$  
      \item EWMA $\tilde{X}$: $\lambda=0.1, L=2.827$  
      \item Ad-CUSUM $\tilde{X}$: $\delta_0=0.15, L=4.344$  
\end{enumerate}
\end{enumerate}

\noindent The parameters were chosen to obtain $ARL_0 \approx 0$. The Table 1 shows the comparison of control charts (a)--(g) on the normally distributed data.  
We can see that, for different values of the shift $\delta$, different types of control charts are the best. For a small shift (0.1), Ad-CUSUM $\tilde{X}$ seems to be best. For medium shifts $0.3--1.0$, EWMA $\bar{X}$ gives the best results; and for the shift $\delta=1.5$, a chart of the Shewhart type is the best.

\medskip
\begin{tabular}{|l|c|c|c|c|c|c|c|}
\hline
Control Chart & \multicolumn{7}{|c|}{shift $\delta$} \\
\cline{2-8}
   & 0.0 &  0.1 &  0.3 &  0.5 &  0.7 &  1.0 &  1.5  \\
\hline
Shewhart $\bar{X}$    &  500.2 &  405.3 &  128.1 &  41.5 &  16.3 &  5.0 &  1.7  \\
\hline
Cusum $\bar{X}$         &  501.2 & 130.0 &  20.2 &  9.9 &  6.5 &  4.5 &  2.9 \\
\hline
EWMA $\bar{X}$         &  501.0 & 136.3 &  19.2 &  8.5 &  5.2 &  3.4 &  2.4 \\
\hline
Shewhart $\tilde{X}$   &  500.1 & 439.2 &  175.3 &  69.6 &  28.3 &  9.7 &  2.7 \\
\hline
Cusum $\tilde{X}$       &  502.0 & 151.4 &  26.2 &  13.1 &  8.7 &  5.7 &  3.8 \\
\hline
EWMA $\tilde{X}$       &  499.2 & 165.4 &  24.4 &  10.3 &  6.4 &  4.0 &  2.7 \\
\hline
Ad-Cusum $\tilde{X}$  &  504.1 & 124.9 &  24.5 &  11.7 &  7.9 &  5.4 &  3.7 \\
\hline
\end{tabular}
\begin{center}Table 1. ARLs of different control charts with data from $N(0,1)$.  
\end{center}

\noindent Table 2 obtained for control charts on the contaminated data is hard to evaluate.  At first glance, EWMA $\bar{X}$ seems to be best for shifts $\delta \geq 0.1$, but ARL(0) is too small and therefore the probability of false alarms is much higher than for the other considered charts.

\medskip
\begin{tabular}{|l|c|c|c|c|c|c|c|}
\hline
Control Chart & \multicolumn{7}{|c|}{shift $\delta$} \\
\cline{2-8}
   & 0.0 &  0.1 &  0.3 &  0.5 &  0.7 &  1.0 &  1.5  \\
\hline
Shewhart $\bar{X}$    &  87.1 &  78.0 &  48.2 &  24.5 &  13.8 &  5.7 &  2.7  \\
\hline
Cusum $\bar{X}$         &  265.4 & 97.4 &  19.2 &  9.9 &  6.6 &  4.4 &  2.9 \\
\hline
EWMA $\bar{X}$         &  186.1 & 85.2 &  17.1 &  8.0 &  5.0 &  3.4 &  2.4 \\
\hline
Shewhart $\tilde{X}$   &  264.5 & 236.8 &  126.6 &  48.6 &  24.1 &  9.8 &  4.0 \\
\hline
Cusum $\tilde{X}$       &  430.0 & 139.4 &  26.8 &  12.8 &  8.6 &  5.8 &  3.8 \\
\hline
EWMA $\tilde{X}$       &  343.9 & 127.4 &  23.3 &  10.0 &  6.2 &  4.0 &  2.7 \\
\hline
Ad-Cusum $\tilde{X}$  &  466.2 & 121.5 &  24.1 &  11.8 &  7.9 &  5.4 &  3.7 \\
\hline
\end{tabular}
\begin{center}Table 2.  ARLs of different control charts with data from a mixture of 94\% $N(\delta, 1)$ and 6\% of $N(\delta, 6.25)$.  
\end{center}

\bigskip

\section{Conclusions}
For comparison purposes, we introduce the $RARL_C$ index, which expresses the relative ARL for different control charts on contaminated data.
 The $RARL_C$ index is defined as follows
 $$
 RARL_C = k.ARL_C(\delta), \ \ \ \text{where} \ \ \  k=\frac{ARL(0)}{ARL_C(0)}.
 $$ 

\noindent Table 3 shows the comparison between different types of control charts using the $RARL_C$ index on the contaminated data.

\medskip
\begin{tabular}{|l|c|c|c|c|c|c|c|}
\hline
Control Chart & \multicolumn{7}{|c|}{shift $\delta$} \\
\cline{2-8}
   & 0.0 &  0.1 &  0.3 &  0.5 &  0.7 &  1.0 &  1.5  \\
\hline
Shewhart $\bar{X}$    &  500 &  447.7 &  276.7 &  140.6 &  79.2 &  32.7 &  15.5  \\
\hline
Cusum $\bar{X}$         &  500 & 183.5 &  36.2 &  18.7 &  12.4 &  8.3 &  5.46 \\
\hline
EWMA $\bar{X}$         &  500 & 228.9 &  45.9 &  21.5 &  13.4 &  9.1 &  6.45 \\
\hline
Shewhart $\tilde{X}$   &  500 & 446.1 &  238.5 &  91.6 &  45.4 &  18.5 &  7.54 \\
\hline
Cusum $\tilde{X}$       &  500 & 162.1 &  31.1 &  14.9 &  10.0 &  6.74 &  4.42 \\
\hline
EWMA $\tilde{X}$       &  500 & 185.2 &  33.9 &  14.5 &  9.01 &  5.82 &  3.93 \\
\hline
Ad-Cusum $\tilde{X}$  &  500 & 130.3 &  25.9 &  12.7 &  8.47 &  5.79 &  3.97 \\
\hline
\end{tabular}
\begin{center}Table 3. Relative ARLs of different control charts with contaminated data.  
\end{center}

\medskip
The last row of the table clearly shows that, in the case of contaminated data, an improved control chart using an adaptive detection scheme is the best choice.

\bigskip\noindent
{\bf Affiliation}: The presented research has been supported by the Ministry of Education of the Czech Republic under the Center of Advanced Aviation Technology CZ.02.1.01/0.0/0.0/16\_019/0000826.


\begin{thebibliography}{9}
\setlength\itemsep{-1mm}

\bibitem{amin} Amin, R. W., Lee, S. J.  (1999).  The effects of autocorrelation and outliers on two-sided tolerance limits. \textit{J. Quality Technology} 31: 286\,--\,300

 \bibitem{baker} Bakir, S. T. (2006). Distribution-Free Quality Control Charts Based on Signed Rank Like Statistics. \textit{Communications in Statistics, Theory and methods}, 35: 
                               734\,--   \,757

\bibitem{capizi} Capizzi, G., Masarotto, G. (2003).  An adaptive exponentially weighted moving average control chart. \textit{Technometrics}, 45: 199\,--\,207.

\bibitem{chakra} Chakraborti, S., Van der Laan, P., Van de Wiel, M. A. (2001) Nonparametric Control Charts: An Overview and Some Results. \textit{J. Quality Technology} 33: 
                               304\,--\,315

\bibitem{davis} Davis R. B., Homer A., Woodall W. H.  (1990). Performance of the zone control chart. \textit{Commun. Statist.-Theory Meth.} 19: 1581\,--\,1587.

\bibitem{dohnal1} Dohnal G. (2009). Design of Control Charts. \textit{ROBUST`08, J\v{C}MF}:  55\,--\,65.

\bibitem{figueir} Figueiredo, F., Gomes, M. I.  (2009). Monitoring industrial processes with robust control charts. \textit{REVSTAT -- Statistical Journal} 7: 151\,--\,170 

\bibitem{jensen} Jensen, W. A., Jones-Farmer, L. A., Champ C. H., Woodall, W. H. (2006). Effects of parameter estimation on a control chart properties: a literature review. 
                                 \textit{J. Quality Technology} 38:  349\,--\,364

\bibitem{lee} Lee, H. Ch., Apley, D. W. (2011). Improved design of robust exponentially weighted moving average control charts for autocorrelated processes. \textit{Quality and Reliability Engineering International} 27: 337\,--\,352

\bibitem{lehman} Lehman, E. L. (1975). \textit{Nonparametrics: Statistical Methods based on Ranks}. Holden-Day. San Francisco, California. 

\bibitem{montgomery} Montgomery, D. C. (2005). \textit{Introduction to Statistical Quality Control}. Wiley, New York.

\bibitem{quesen} Quesenberry, D. C. (1993). The effect of sample size on estimated limits for X-bar and X control charts. \textit{J. Quality Technology} 25: 206\,--\,247 

\bibitem{reynolds} Reynolds, M. R., Stoumbos, Z. G.  (2010). Robust CUSUM charts for monitoring the process mean and variance. \textit{Quality and Reliability Engineering International} 26:  453\,--\,473

\bibitem{rocke} Rocke, D. M.  (1989). Robust control charts. \textit{Technometrics}, 31:  173\,--\,184

\bibitem{sparks} Sparks, R. S. (2000). CUSUM charts for signaling varying location shifts. \textit{J. Quality Technology}, 32:  157\,--\,171.

\bibitem{tagaras} Tagaras, G. (1998).  A survey of recent developments in the design of adaptive control charts. \textit{J. Quality Technology} 30: 212-231

 \bibitem{tsung} Tsung, F., Wang, K. (2010).  Adaptive Charting Techniques: Literature Review and Extensions. \textit{Frontiers in Statistical Quality Control} 9, Springer Physica Verlag, Heidelberg

\bibitem{woodal} Woodall, W. H., Montgomery, D. C. (1999).  Research issues and ideas in statistical process control. \textit{J. Quality Technology} 31: 376\,--\,386


\bibitem{yang}Yang L., Pai S., Wang Y. R. (2010).  A novel CUSUM Median Control Chart. \textit{Proceedings of the International MultiConference of Engineers and Computer Scientists 2010} Vol. III, IMECS 2010, March 2010, Hong Kong, 17\,--\,19

\bibitem{zimmer} Zimmer, L.S., Montgomery D.C., Runger G.C. (2000). Guidelines for the application of adaptive control charting schemes. \textit{International Journal of Production Research} 38: 1997\,--\,1992
\end{thebibliography}
\end{document}